# Wafer-scale High-*k* SrTiO$_3$ Dielectrics with Rational Barrier-layer Design for Low Leakage and High Charge Density

Majid Mohseni[*], Shivasheesh Varshney, Seung Gyo Jeong, Amber Walton, C. Daniel Frisbie, and Bharat Jalan[*]

Department of Chemical Engineering and Materials Science, University of Minnesota, Twin Cities, Minneapolis, MN 55455

[*]Corresponding authors: mohse012@umn.edu; bjalan@umn.edu

**ABSTRACT**

High-$k$ oxides such as SrTiO$_3$ promise large capacitance, but their dielectric response is often limited by leakage currents due to reduced bandgaps. We show that introducing a thin barrier layer beneath SrTiO$_3$ is a simple and effective way to suppress leakage and increase charge density. Using hybrid molecular beam epitaxy, we grew uniform SrTiO$_3$ films on Nb:SrTiO$_3$, CaSnO$_3$/Nb:SrTiO$_3$, and 2-inch SiO$_2$/p-Si stacks to directly compare how different barrier layers influence device behavior. Both CaSnO$_3$ and SiO$_2$ reduce leakage, but the ultra-wide-bandgap SiO$_2$ layer enables much higher operating voltages, yielding charge densities exceeding $5 \times 10^{13}$ cm$^{-2}$ at room-temperature – more than a fivefold enhancement compared to devices without a barrier layer. This improvement comes with a predictable trade-off: the lower dielectric constant of SiO$_2$ reduces overall capacitance, making its thickness an important design parameter. Together, these results demonstrate that rational barrier-layer engineering – including wafer-scale integration on Si – provides a clear pathway to achieving higher charge densities in SrTiO$_3$-based dielectric devices.

**Keywords:** Dielectrics, high-$k$, perovskite oxides, hybrid MBE, dielectric constants, leakage

INTRODUCTION

The amount of charge stored in a dielectric is set by the simple relation Q = $\int C dV$, where C is the capacitance, and *V* is the applied voltage.[1] Achieving large Q therefore requires two conditions: (i) a high capacitance, which scales with the dielectric constant (*k*) and inversely with thickness, and (ii) a voltage window wide enough to support large fields without breakdown or leakage. In practical dielectric stacks, these requirements are often in conflict. Materials with high-*k* typically have smaller bandgaps ($E_g$) (Figure 1), which reduce the energy barrier for electron injection at the metal-dielectric interface resulting in higher leakage.[2] Leakage currents therefore limit the maximum usable voltage well before the intrinsic polarization response of the dielectric is reached. This trade-off can be illustrated by comparing two widely used classes of dielectrics. Ultra-wide-bandgap insulators such as $SiO_2$ ($E_g$ = 9 eV, *k* = 3.9)[3] suppress leakage exceptionally well and tolerate electric fields up to 10 MV/cm,[4] but their small *k* yields modest capacitance. In contrast, perovskite oxides such as $SrTiO_3$ (STO) and $BaTiO_3$ (BTO) exhibit very large dielectric constants – STO has *k* = ~ 300 at room-temperature[5] (increasing to ~ 25000 at 2 K),[5] while BTO exceeds *k* = 3000 near room-temperature[6] – but their bandgaps of ~ 3.2 eV[7,8] produce smaller band offsets with metal electrodes.[9] This lowers the interface barrier height and restricts the voltage that can be applied. Thus, the dielectric response of high-*k* oxides is typically limited not by their intrinsic dielectric constants but by interface-controlled leakage.

Introducing a thin, wide-bandgap barrier layer between a high-*k* oxide and metal electrode offers a quantitative route to overcome this limitation. A barrier layer increases the conduction-band offset, raises the tunneling barrier height, and redistributes the electric field across the stack. Even a few nanometers of a material such as ultra-wide-bandgap $SiO_2$ or a wide-gap perovskite like $CaSnO_3$ (CSO, $E_g$ = 4.2 – 4.9 eV)[10] can dramatically reduce leakage current and extend the accessible voltage range, while the high-*k* layer (e.g., STO) continues to dominate the capacitance. The thickness and permittivity of the wide-bandgap barrier layer therefore become critical design parameters: too thin, and leakage dominates; too thick, and the overall capacitance is reduced.

In this work, we investigate how the bandgap-driven barrier height controls achievable charge density in dielectric stacks. Using hybrid molecular beam epitaxy, we fabricate two architectures with precisely controlled thicknesses: fully epitaxial 5 mm × 5 mm STO/CSO/Nb:STO (001) heterostructures and wafer-scale (2") STO/SiO$_2$/p-Si (001) stacks. CSO provides a moderate bandgap increase over STO (4.2 - 4.9 eV vs. 3.2 eV), while SiO$_2$ has an ultra-wide-gap (~9 eV). By systematically comparing capacitance, leakage currents, and maximum stored charge, we show that increasing the interface barrier height directly increases the usable voltage window and enables charge densities above $5 \times 10^{13}$ cm$^{-2}$ in STO-based dielectrics, which is more than a fivefold enhancement compared to devices without a barrier layer. These results demonstrate that rational barrier-layer engineering – rather than relying on a single "ideal" dielectric – offers a clear and quantitative pathway to higher charge-density operation in oxide-based devices.

RESULTS AND DISCUSSION

To establish a baseline for the charge-density limits of STO, we first grew STO films directly on conductive 0.5 wt.% Nb-doped STO (001) substrates (5 mm × 5 mm) using hybrid MBE.[11] The films were annealed in oxygen at 950 °C for 2 hours to minimize oxygen vacancy concentrations. Figure 2a summarizes the structural and surface quality of a representative 300 nm STO film. High resolution X-ray diffraction (HRXRD) and reflection high energy electron diffraction (RHEED) confirm that the film is fully epitaxial and single crystalline, and the atomic force microscopy (AFM) image shows an atomically smooth surface with a root mean square (RMS) roughness of 0.36 nm. To measure the dielectric properties, Pt top electrodes were deposited by sputtering onto photolithographically patterned pads of different sizes (50 to 500 μm diameter), with the Nb:STO substrate serving as the bottom electrode. Impedance measurements were performed using an impedance analyzer. As shown in Figure 2b, the 300 nm STO film exhibits a phase angle of nearly -90° across the 20 - 1000 Hz range, indicating nearly ideal capacitor behavior. From the measured impedance, we extracted a dielectric constant of 291 – consistent with the reported bulk value of ~300.[12] STO films with a thickness of 100 nm showed similar structural quality and dielectric response, confirming reproducibility of the growth process (Figure S1). Because achievable charge density depends not only on capacitance but also on the voltage range that can be applied without significant leakage (e.g. < $10^{-6}$ A/cm$^2$), we performed current-voltage (I-V) measurements. Figure 2c shows the resulting voltage windows for two different thicknesses of STO films, defined using a leakage current limit of $10^{-6}$ A/cm$^2$. For 100 nm STO, the usable voltage range is narrow (- 1.02 V to + 0.34 V) whereas the 300 nm STO film shows a larger negative-bias window (down to - 4.51 V). In all measurements, the bias was applied to the top electrodes while the bottom electrodes were held at ground. For the 300 nm STO film, the corresponding maximum charge densities are approximately $2.15 \times 10^{13}$ cm$^{-2}$ (for negative bias) and $1.83 \times 10^{12}$ cm$^{-2}$ (for positive bias), respectively. The corresponding charge densities for a 100 nm STO film are $1.42 \times 10^{13}$ cm$^{-2}$ (for negative bias) and $4.73 \times 10^{12}$ cm$^{-2}$ (for positive bias). These results show that simply decreasing STO thickness does not improve charge density: thinner films offer higher capacitance but break down at lower voltages, while thicker films tolerate higher voltages but store less charge per unit area. These results highlight the need for an approach that expands the voltage window in both polarities to enable higher overall charge density.

To address this limitation, we incorporated an epitaxial CSO barrier layer because of its wider bandgap (~ 4.7 eV)[10] and a lattice parameter (3.95 Å)[13,14] that closely matches that of STO. Figure 2d shows HRXRD data from a 100 nm STO (001) film grown on an 8 nm

CSO/Nb:STO (001) layer confirming epitaxial, single crystalline film growth. RHEED patterns taken after growth further verify the crystalline quality and epitaxial growth of both the CSO (001) and STO (001) layers. Like the STO-only samples, the heterostructure was annealed in oxygen at same conditions to minimize oxygen vacancies. AFM imaging confirms an atomically smooth surface, with an RMS roughness of 0.14 nm (inset of Figure 2d). Figure 2e shows the frequency-dependent phase angle and impedance magnitude, exhibiting a phase angle of - 90° across the 10 - $10^5$ Hz range which is larger than the STO-only samples. From these measurements, the resulting $k$ ($k_{effective}$) was determined to be 202 assuming $C_{measured} = \frac{k_{effective} \times A \times \varepsilon_0}{t_T}$, where $C_{measured}$ is the measured capacitance value, A is the pad area, $\varepsilon_0$ is the permittivity of free space, and $t_T$ is the total thickness of dielectric layers (here is 108 nm). This $k_{effective}$ value is lower than that of bulk STO and our STO-only film, but it is significantly higher than that of bulk CSO, as expected from the series combination of the 100 nm STO layer and the 8 nm CSO barrier. Leakage currents were evaluated using I-V measurements. As shown in Figure 2f, the CSO barrier significantly widens the usable voltage window to - 2.2 V to + 0.8 V, a substantial improvement over 100 nm STO grown directly on Nb:STO. This wider operating range enables higher charge storage, corresponding to a charge density of $2.3 \times 10^{13}$ cm$^{-2}$ (for negative voltage) and $8.25 \times 10^{12}$ cm$^{-2}$ (for positive voltages). These results demonstrate a nearly twofold enhancement in charge density by incorporating a wide-bandgap CSO barrier layer despite losing on the total capacitance.

Building on these results, we next explored SiO$_2$ as a barrier layer. SiO$_2$ has an ultra-wide bandgap ~ 9 eV and forms naturally on silicon, a low-cost, widely available substrate that can be manufactured in large wafer sizes. STO films with thicknesses ranging from 40 to 125 nm were grown on 2-inch p-type Si wafers by hybrid MBE, with the p-Si substrate functioning as the bottom electrode. Figure 3a provides an overview of the STO/Si architecture – including a schematic, optical image and ellipsometry thickness map of 125 nm sample. The SEM of a 125 nm sample, and AFM images and RHEED patterns of different thicknesses of STO are included in Figure S2. The RHEED patterns indicate polycrystalline STO, as expected from the growth on an amorphous $SiO_2$ interlayer at the interface. The SEM image shows a uniform surface morphology, and ellipsometry mapping reveals excellent thickness uniformity with < 2.5% variation across the 2" wafer. AFM images for multiple STO thicknesses show consistently smooth surfaces on the nanometer scale. These results along with the optical conductivity spectra which were collected at multiple regions across the 2" wafer (Figure S3) confirm that STO layers can be grown uniformly on large-area Si wafers.

To assess dielectric uniformity across the 2-inch wafer, the 125 nm STO/$SiO_2$(~ 6 nm)/Si (100) sample was diced into ~ 7 mm × 7 mm chips, and nine regions spanning the wafer were selected for electrical testing. Pt top electrodes were deposited on each chip using identical lithographically defined patterns to ensure consistent electrode geometry. Impedance measurements were performed using an impedance analyzer. Figure 3b shows the magnitude of impedance and the phase angle as a function of frequency from region 9. This data exhibits ideal capacitor-like behavior. An effective dielectric constant of 52 was extracted – lower than that of STO-only devices, yet significantly higher than that of conventional $SiO_2$ devices, due to the capacitive series combination of STO with $SiO_2$. Note that the $SiO_2$ thickness required to calculate the effective dielectric constant is obtained from Figure S4, which provides the total thickness of the dielectric stack. Dielectric constants measured at all nine regions (Figure 3c) show minimal variation, demonstrating excellent uniformity across the full wafer. As shown in Figure 3d, all samples exhibited consistent leakage characteristics, with voltage limits ranging from approximately - 5.5 V to + 1.5 V (corresponding I-V curves shown in Figure S5). These results confirm that $SiO_2$ enables a significantly wider voltage window without introducing substantial leakage. Importantly, these STO/$SiO_2$/Si samples were not annealed because high-temperature oxygen annealing, while reducing oxygen vacancies in STO, would also grow additional SiO$_2$ by oxidizing the silicon substrate. This trade-off may increase the voltage window but can decrease capacitance, ultimately lowering charge density. Optimizing the annealing process is therefore critical for maximizing performance in STO/SiO$_2$ devices (see SI Figure S6). C-V data at 1 kHz (Figure 3e) show stable capacitance across the tested voltage range, and a linear dependence of capacitance on electrode area for pad diameters of 100-500 μm confirms that the extracted dielectric constant is geometry-independent.

Figures 4a and 4b summarize how capacitance and voltage window change after annealing (30 min at 950°C) for different STO thicknesses. A key result is that the breakdown limits hardly change with STO thickness. Instead, they are controlled almost entirely by the SiO$_2$ layer. For instance, the device with 40 nm STO shows a very wide operating window from -19 V to +7 V (corresponding I-V curves are provided in Figure S7). The calculated SiO$_2$ thickness for the optimized condition is in range of 7 to 9 nm for different thicknesses of STO based on Figure S4. This value can increase to over 30 nm if we anneal the samples more than 2 hours at 950°C. Across all samples, the 40 nm STO device achieves the highest charge density because it strikes the best balance between having high capacitance and being able to withstand a large voltage range (Figure 4c). It reaches a charge density of $5.25 \times 10^{13}$ cm$^{-2}$ for negative bias and a charge density of $2 \times 10^{13}$ cm$^{-2}$ for positive bias. All values were measured at 1 kHz to account for the frequency dependence of the dielectric response. The complete frequency-dependent behavior of capacitance and effective dielectric constant for this device is shown in Figure S8 which displays the expected drop in capacitance at higher frequencies.

Figure 4d compares the charge density per unit area against the dielectric constant for a range of dielectric materials, including the STO/SiO$_2$/Si structure developed in this work. Our device performs on par with, and in many cases exceeds, leading reports in the field. When combined with the low cost of silicon, the availability of large wafer sizes, this platform provides a practical and scalable route for building large-area devices requiring high charge density/area.

CONCLUSION

This work demonstrates that controlling the dielectric stack is essential for achieving high charge densities. While STO provides a high dielectric constant, its usable voltage range is limited by leakage. Introducing wide-bandgap barrier layers – first CSO and then $SiO_2$ – significantly expands the voltage window and enables higher charge storage. In particular, the $STO/SiO_2/Si$ platform combines large charge densities, uniform large-area growth, and the scalability of silicon wafers, making it a strong candidate for practical, wafer-scale devices.

**MATERIALS AND METHODS**

Hybrid molecular beam epitaxy was employed to grow STO and CSO thin films. STO layers were grown on single crystalline 0.5 wt. % Nb-doped $SrTiO_3$ (001) substrates (Crystec GmbH) and 2-inch p-type silicon wafers (MTI Corporation) using identical growth procedures. The base pressure of the MBE chamber (Scienta Omicron Inc) was $1 \times 10^{-8}$ Torr. Prior to STO growth on silicon, wafers underwent oxygen plasma cleaning (250 W, $8 \times 10^{-6}$ Torr $O_2$ flow, 2 minutes) to remove surface contaminants and ensure a clean growth interface. During STO growth, strontium was supplied from a titanium crucible (MBE Komponenten, Inc.) loaded with 99.99% pure Sr (Sigma-Aldrich), maintained at a beam equivalent pressure (BEP) of $9.07 \times 10^{-8}$ Torr. Titanium was introduced via titanium isopropoxide (TTIP, 99.999%, Sigma-Aldrich) delivered through a gas inlet system at a baratron pressure of 165 mTorr, with the oxygen flow rate matched to the plasma cleaning step. The substrate temperature during STO growth was held at 900 °C, and films were cooled under an oxygen plasma environment to minimize oxygen vacancies. For CSO growth, calcium was evaporated from a titanium crucible containing 99.99% pure Ca (Sigma-Aldrich) at a BEP of $6.15 \times 10^{-9}$ Torr. Tin was supplied using hexamethyl ditin (HMDT, 99%, Sigma-Aldrich) through a gas inlet system at a baratron pressure of 260 mTorr, corresponding to a BEP of $1.02 \times 10^{-5}$ Torr. The substrate temperature was maintained at 950 °C during deposition. Oxygen plasma (250 W, $5 \times 10^{-6}$ Torr $O_2$ flow) applied during the growth.

Reflection high-energy electron diffraction (RHEED, Staib Instruments) was employed in situ during and after film growth to monitor surface crystallinity and morphology in real time. Post-growth surface characterization was performed using atomic force microscopy (AFM, Bruker) to assess surface roughness and topography. High-resolution X-ray diffraction (HRXRD) measurements were conducted using a SmartLab XE diffractometer (Rigaku) to evaluate the structural quality, phase purity, and epitaxy of the deposited films.

The device fabrication process began with the application of a negative photoresist (NR71-3000P) onto the STO surface, which was uniformly coated using a spin coater (Apogee™ Spin Coater). To define the electrode pattern, photolithography (SUSS MicroTec MA/BA6) was employed. The sample was aligned with a photomask and then subjected to UV exposure for 22 seconds, crosslinking the exposed regions of the resist. After exposure, the unexposed areas were removed by immersion in developer (RD6), revealing the underlying STO in the desired pattern. A 100 nm-thick platinum layer was then deposited using sputtering (AJA International, Inc.), producing uniform and adherent metal films. Finally, a lift-off process was conducted using photoresist remover (RR-41) to dissolve the remaining resist along with the overlying platinum, leaving clean and sharply defined Pt features only in the patterned areas.

Dielectric measurements were performed using a semiconductor parameter analyzer (Keysight B1500A) and an impedance analyzer (Keysight Technologies), both operated with a probe station for precise electrical probing. Film thickness and uniformity across the 2-inch wafers were confirmed using a spectroscopic ellipsometer (Rudolph Technologies, Inc.). After

STO growth, the wafers were sectioned into smaller pieces using a precision dicing saw (DISCO Corporation) to enable localized electrical and structural characterization. The optical conductivity spectra of STO/SiO$_2$/Si sample were investigated using spectroscopic ellipsometers (VASE, J. A. Woollam Co.) at room temperature. The optical spectra were obtained between 2 - 5 eV for incident angles of 65, 70 and 75° and estimated by using a two-layer model (film and substrate).

## ASSOCIATED CONTENT

### Data Availability Content

All data needed to evaluate the conclusions of the paper are present in the paper and/or the Supporting Information.

### Supporting Information

The Supporting Information is available free of charge.

The supporting information includes the structural and dielectric characterization of 100 nm STO/Nb:STO (001) sample, structural characterization of STO grown on 2-inch Si wafers, Position-dependent optical spectra measured by spectroscopic ellipsometry for 60 nm STO sample grown on 2-inch Si wafer, Calculated thickness of silicon dioxide for different STO thicknesses, IV characteristics for 9 different regions of 125 nm STO on 2-inch Si wafer, Charge density optimization by oxygen annealing, IV Characteristics for different thicknesses of STO on Si after annealing at optimized condition, and Frequency-dependent measurements of capacitance and effective dielectric constant for 40 nm STO on Si sample.

## AUTHOR INFORMATION

Corresponding authors

Bharat Jalan - Department of Chemical Engineering and Materials Science, University of Minnesota, Twin Cities, Minneapolis, MN 55455. Email: Bjalan@umn.edu

Majid Mohseni - Department of Chemical Engineering and Materials Science, University of Minnesota, Twin Cities, Minneapolis, MN 55455. Email: mohse012@umn.edu

Authors

Shivasheesh Varshney - Department of Chemical Engineering and Materials Science, University of Minnesota, Twin Cities, Minneapolis, MN 55455

Seung Gyo Jeong - Department of Chemical Engineering and Materials Science, University of Minnesota, Twin Cities, Minneapolis, MN 55455

Amber Walton - Department of Chemical Engineering and Materials Science, University of Minnesota, Twin Cities, Minneapolis, MN 55455

C. Daniel Frisbie - Department of Chemical Engineering and Materials Science, University of Minnesota, Twin Cities, Minneapolis, MN 55455

**ACKNOWLEDGEMENTS**

Synthesis and characterization (M.M., and S.V.) were supported by the Center for Programmable Energy Catalysis, an Energy Frontier Research Center funded by the U.S. Department of Energy, Office of Science, Basic Energy Sciences at the University of Minnesota, under Award No. DE-SC0023464, and partly through DE-SC0020211. Film growth was performed using instrumentation funded by AFOSR DURIP award FA9550-18-1-0294. Parts of this work were carried out at the Characterization Facility, University of Minnesota, which receives partial support from the NSF through the MRSEC program under award DMR-2011401. Exfoliation of films and device fabrication was carried out at the Minnesota Nano Center, which is supported by the NSF through the National Nano Coordinated Infrastructure under award ECCS-2025124.

**AUTHORS CONTRIBUTION**

M.M., S.V., C.D.F. and B.J. conceived the idea and designed the experiments. M.M. and S.V. grew the films. M.M. and S.V. characterized the films. M.M. fabricated the devices. M.M and A.W. performed the electrical measurements. S.G.J. measured optical conductivity spectra. M.M, S.V., and B.J. wrote the manuscript. All authors contributed to the discussion and manuscript preparation. B.J. directed the overall aspects of the project.

Heterostructure Field Effect Transistors. *IEEE Electron Device Lett.* **2020**, *41* (4), 621–624. https://doi.org/10.1109/LED.2020.2976456.

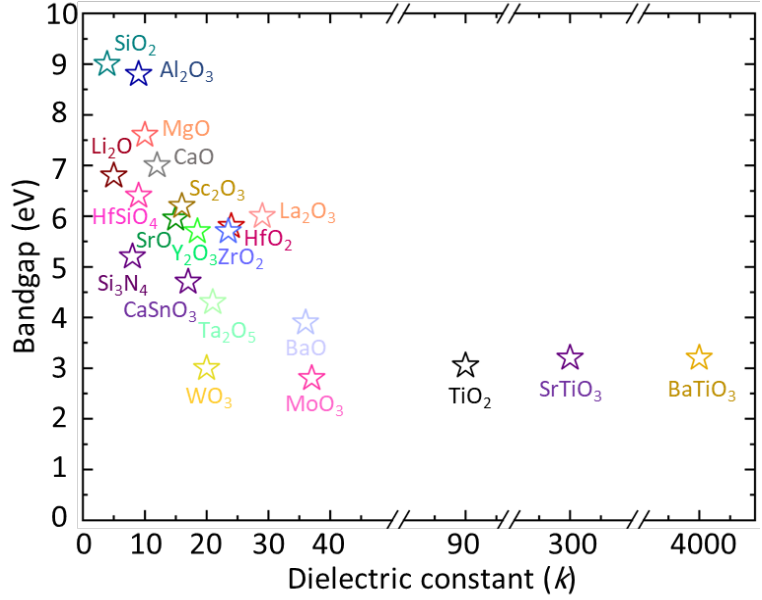

**Figure 1:** Bandgap ($E_g$) as a function of dielectric constant (k) for $SiO_2$,[2,15] $Al_2O_3$,[15,16] $Li_2O$,[15] $MgO$,[2] $CaO$,[2] $HfSiO_4$,[15] $Sc_2O_3$,[15] $SrO$,[15] $Si_3N_4$,[16] $CaSnO_3$,[10,17] $Y_2O_3$,[2] $La_2O_3$,[16] $HfO_2$,[2] $ZrO_2$,[2,16] $Ta_2O_5$,[2,15,16] $WO_3$,[15] $BaO$,[2,15] $MoO_3$,[15] $TiO_2$,[18] $SrTiO_3$,[5,7] $BaTiO_3$.[8,19] Dielectrics with higher dielectric constant typically have lower bandgaps.

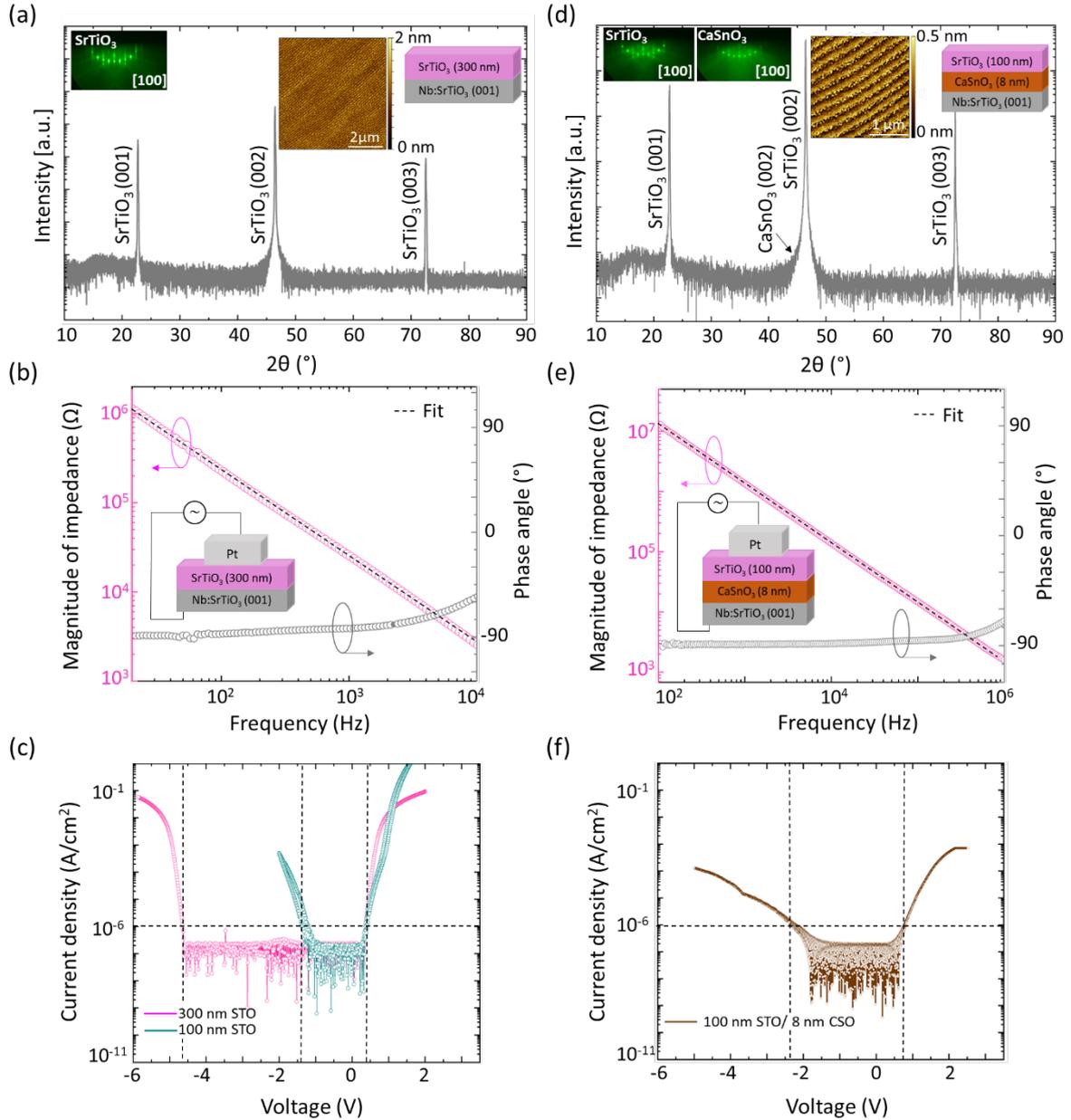

**Figure 2:** Structural and dielectric characterization of STO grown on Nb:STO (001) and on a CSO barrier layer (a) 2θ-ω coupled X-ray diffraction scan of 300 nm STO grown on Nb-doped STO (001) substrate via hybrid MBE. Insets display the AFM image (rms roughness of 0.36 nm), RHEED pattern, and schematic diagram of the STO/Nb:STO structure. (b) Frequency-dependent impedance magnitude and phase angle of the Pt/STO (300 nm)/Nb:STO capacitor, demonstrating ideal dielectric behavior. The extracted dielectric constant for this sample is 291. (c) Current density versus voltage (J-V) characteristics for 100 nm and 300 nm Pt/STO/Nb:STO devices, measured with a positive and a negative voltage applied to the top electrode. (d) 2θ-ω coupled X-ray diffraction scan of a 100 nm STO film deposited on an 8 nm CSO barrier layer grown on a Nb-doped STO (001) substrate via hybrid MBE. Insets show RHEED images of the CSO layer and the overlying STO film, along with an AFM image of STO (rms roughness of 0.14 nm) and schematic of the STO/CSO/Nb:STO heterostructure. (e) Impedance magnitude and phase angle as a function of frequency for the Pt/ STO (100 nm)/CSO (8 nm)/Nb:STO capacitor. The calculated effective dielectric constant is about 202 for this sample. (f) Current density-voltage (J-V) characteristics of the Pt/STO (100 nm)/CSO (8 nm)/Nb:STO structure, highlighting the extended voltage window enabled by the CSO barrier layer.

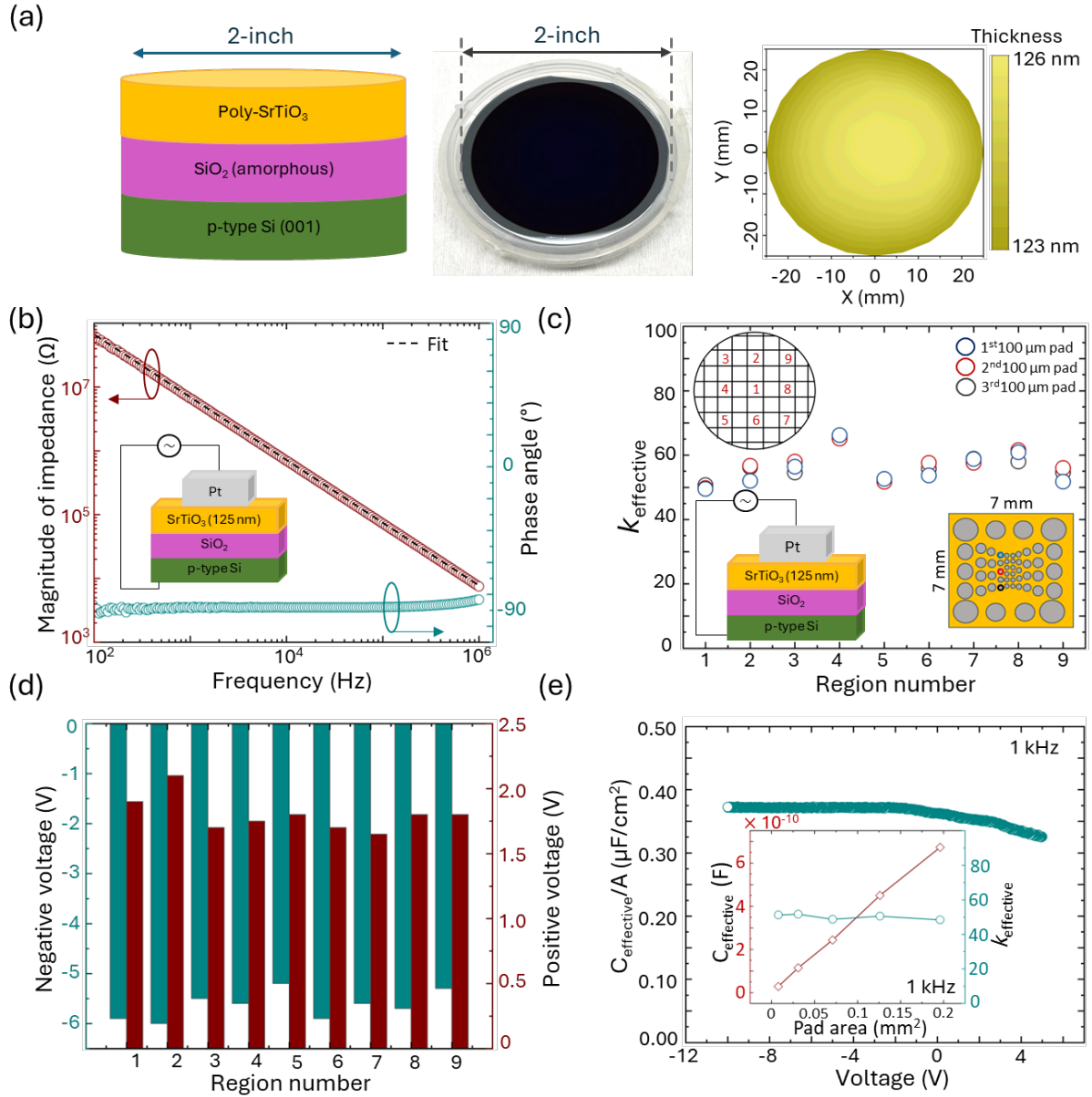

**Figure 3.** Uniform dielectric properties of STO across a 2-inch silicon wafer. (a) Sample schematic, optical image, and ellipsometry thickness map of 125 nm STO grown by hybrid MBE on a 2-inch p-type silicon wafer. (b) Impedance magnitude and phase angle as a function of frequency for the Pt/STO (125 nm)/SiO$_2$/p-Si capacitor, showing ideal dielectric behavior with extracted dielectric constant of about 52. (c) Effective dielectric constant of 125 nm STO measured at 9 different regions (3 different pads of 100 μm diameter for each region) across the 2-inch wafer. Insets show the mapped measurement locations, device schematic, and Pt pad patterns (pad sizes from 50 to 500 μm diameter) on STO. (d) Voltage window (defined at a leakage current threshold of $10^{-6}$ A/cm$^2$; positive and negative limits shown separately) for the same 9 regions as in (c), confirming spatial uniformity in breakdown behavior across the wafer. (e) Capacitance per unit area (C/A) of 125 nm STO sample as a function of applied voltage at 1 kHz, showing the voltage-independent behavior of capacitance. The inset shows the extracted effective dielectric constant and capacitance as a function of top Pt pad area, confirming area-independent behavior.

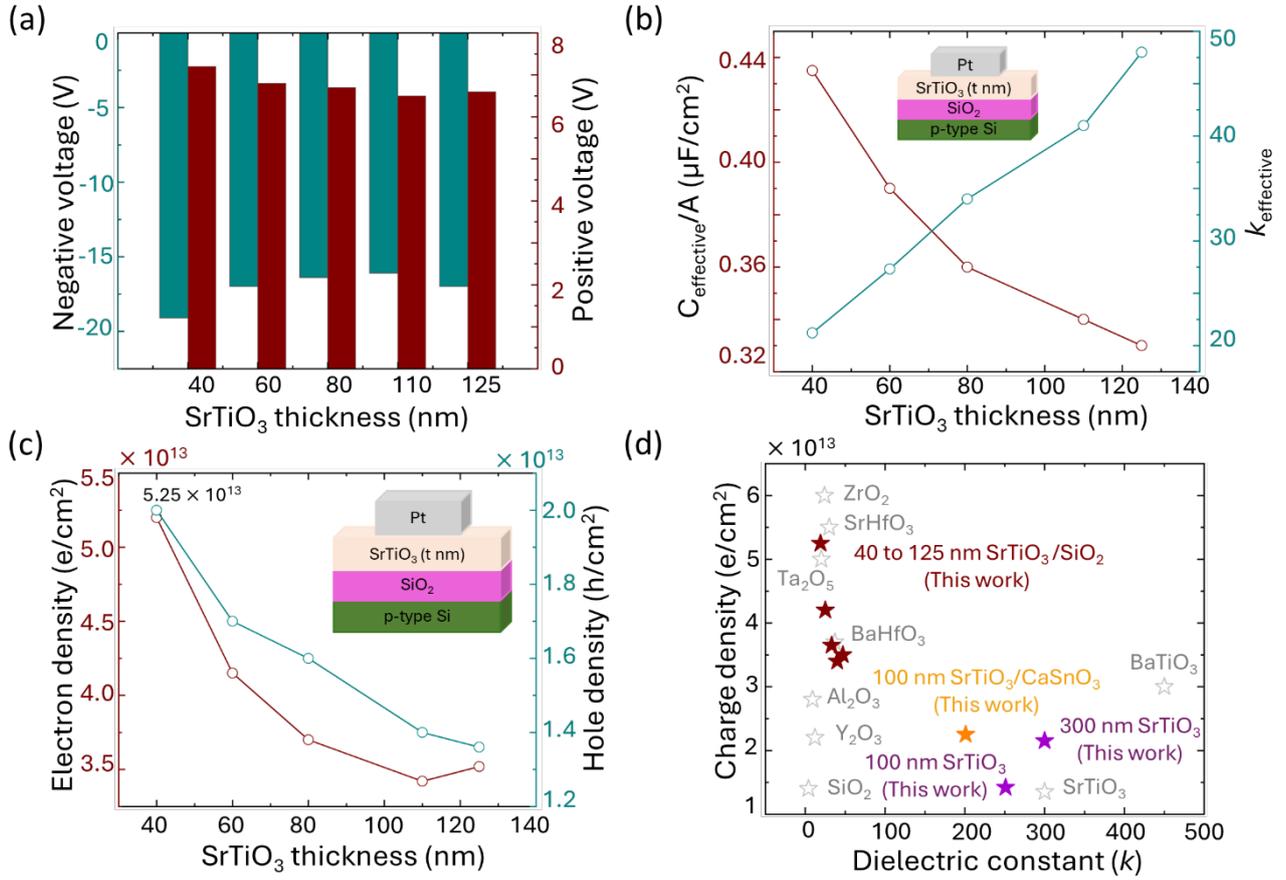

**Figure 4**. Charge density calculation and optimization. (a) Voltage window (defined at a leakage current threshold of $10^{-6}$ A/cm$^2$) for Pt/STO/SiO$_2$/p-Si devices with varying STO thicknesses after annealing under optimized conditions. (b) The device schematic, and the effective dielectric constant and capacitance per unit area as functions of STO thickness after optimized annealing condition. (c) Optimized electron and hole densities for different STO thicknesses in Pt/STO/SiO$_2$/p-Si devices. (d) Charge density comparison of the hybrid MBE-grown STO on different barrier layers reported in this study with other dielectric materials reported in the literature[20] including polycrystalline ZrO$_2$,[20–22] single crystalline SrHfO$_3$,[20,23] amorphous Ta$_2$O$_5$,[20,24,25] single crystalline BaHfO$_3$,[20,26] single crystalline Al$_2$O$_3$,[20,25,27] single crystalline Y$_2$O$_3$,[20,27] amorphous SiO$_2$,[20,25] single crystalline SrTiO$_3$,[20,25,28] and single crystalline BaTiO$_3$.[20,29,30]